\documentclass{article}
\usepackage[T1]{fontenc}
\usepackage{spconf,amsmath,amssymb,url,booktabs,graphicx}
\usepackage[hidelinks]{hyperref}
\usepackage{orcidlink}

\title{SEER: Source-Conditioned Emotion Enhancement\\
via Retrieval for Cochlear-Implant Speech}
\name{Hsing-Hang Chou\orcidlink{0009-0005-9806-2174}$^{1}$ \quad
Yun-Shao Lin\orcidlink{0000-0002-1494-1413}$^{1}$ \quad
Ching-Chin Sung\orcidlink{0009-0005-6774-480X}$^{2}$ \quad
Chi-Chun Lee\orcidlink{0000-0003-0186-4321}$^{1}$}
\address{$^{1}$Department of Electrical Engineering, National Tsing Hua University, Taiwan\\
$^{2}$Research Center for Information Technology Innovation, Academia Sinica, Taiwan}

\begin{document}
\maketitle

\begin{abstract}
Cochlear implants (CIs) restore speech access but weaken cues needed for vocal emotion recognition. Prior CI-oriented enhancement requires parallel normal/strong recordings and intensity labels. We propose SEER, a retrieval-based framework that learns which same-emotion reference helps each source remain recognizable after CI processing. A source-conditioned retriever learns CI-aware utility from sampled emotional voice conversion outcomes, while uncertainty-guided exploration avoids exhaustive pair evaluation; neither parallel recordings nor intensity labels are required. SEER improves Source macro-F1 at N8 by 7.30 points on RAVDESS and 11.66 points on ESD, with significant ESD gains across N4/N8/N16. Sixteen-listener RAVDESS gains are significant across all conditions. Exhaustive analysis finds an aggregate benefit from stronger references but little effect from matching gender or content.
\end{abstract}


\begin{figure*}[t]
\centering
\includegraphics[width=\textwidth,height=50mm]{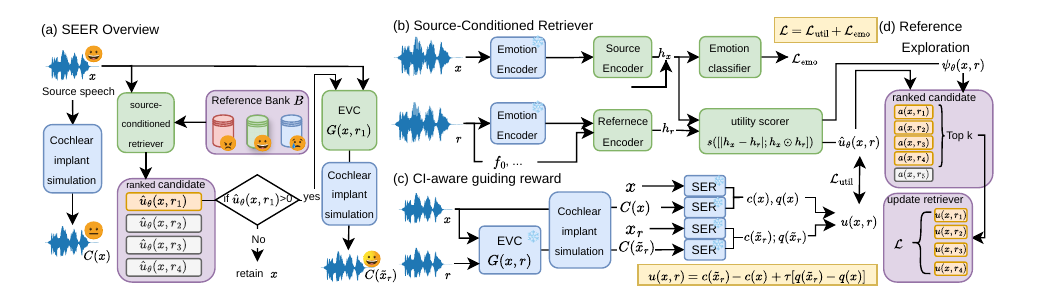}
\caption{SEER framework: (a) retrieves a reference to improve post-CI emotion recognizability; (b) estimates source--reference utility; (c) derives its CI-aware target; and (d) uses acquisition scores to sample pairs for retriever updates.}
\label{fig:pipeline}
\end{figure*}

\section{Introduction}

Cochlear implants (CIs) restore auditory access to speech, and conventional CI outcome assessment has long centered on spoken-word and sentence recognition~\cite{shafiro2025evolving}. Effective communication, however, depends not only on linguistic content but also on emotional information conveyed through speech. Emotion perception has received comparatively less attention in CI research~\cite{sahana2024vocal}. This gap is particularly important because electric hearing inadequately transmits fundamental-frequency cues that contribute strongly to emotion perception~\cite{everhardt2020}. Consistent with this limitation, both CI users and normal-hearing listeners presented with CI-vocoded speech show reduced emotion-recognition accuracy, with CI users also tending toward Neutral responses~\cite{everhardt2020,luo2007}. Recent work evaluating speech emotion recognition (SER) models under CI simulation similarly reports degradation in emotion-recognition performance~\cite{mobassara2025a}.

Prior perceptual studies suggest that some emotional expressions remain more recognizable than others after CI processing. Exaggerated child-directed emotional speech has shown higher recognition than adult-directed speech under CI-simulated spectral degradation~\cite{xie2026spectral}, while similar benefits of exaggerated emotional prosody and emotion caricatures have been observed in CI users~\cite{barrett2020perception,voneiff2022}. Building on such findings, Brueggeman and Hansen~\cite{brueggeman2022} used JD-NMF~\cite{fu2017jdnmf} to learn normal-to-strong emotion conversion from paired utterances of the same speaker and sentence. Recorded strong speech was more recognizable overall, whereas converted speech yielded limited, emotion-dependent listener gains. More importantly, this approach depends on both explicit normal/strong intensity distinctions and parallel source--target recordings, limiting its applicability to broader emotion datasets that lack such supervision or matched utterance pairs.

Recent reference-conditioned emotional voice conversion (EVC) removes this paired-data requirement by transferring emotional expression from an independent reference while preserving the source content and speaker characteristics~\cite{chou2025zsdevc,maestro2025}. This makes ordinary categorical emotion corpora usable as a reference bank for enhancement, rather than requiring matched normal/strong recordings. For CI emotion enhancement, the remaining problem is therefore not how to obtain a parallel strong target, but which emotional template should guide each source. Because EVC combines a reference expression with a different source speaker and utterance, cues that make the reference recognizable after CI processing may not transfer as effectively to the converted speech. Exhaustively evaluating every source--reference pair could identify suitable templates, but the cost grows rapidly with corpus and reference-bank size. A scalable enhancement pipeline therefore needs to learn source-specific reference selection from only a subset of these post-conversion outcomes.

We therefore propose SEER, a source-conditioned emotion enhancer that learns to retrieve an effective reference for each source. Its retriever predicts conversion utility and is guided by a CI-aware reward derived from SER evaluation of EVC outputs after CI processing. Acquisition-based exploration balances this prediction with uncertainty, preventing sampled targets from concentrating on current favorites. Together, they train reference selection without parallel recordings or intensity labels. Experiments evaluate emotion retention, intelligibility, and whether intensity, gender, and content explain reference utility.

\section{Methodology}
\label{sec:method}

SEER uses reference-conditioned EVC to enhance each source toward a same-emotion reference, retaining the source when no candidate has positive predicted utility. Recordings recognized correctly before and after CI simulation form the reference bank. Because exhaustive source--reference conversion does not scale, the retriever and exploration rule sample pairs whose CI-aware outcomes guide learning. Figure~\ref{fig:pipeline} summarizes the framework.

\subsection{Reference-Conditioned Emotion Enhancement}
\label{sec:evc-enhancement}

To realize this enhancement, we instantiate converter $G$ with Maestro-EVC~\cite{maestro2025}. Given source $x$ expressing $e_x$ and reference $r$ expressing $e_r$, $G(x,r)$ produces $\tilde{x}_r$ by shifting the source toward the reference expression while preserving linguistic content and speaker identity. Each training recording thus defines a possible enhancement direction without requiring a parallel pair.

To identify suitable references, we use pretrained Emotion2Vec+ Large~\cite{ma2024emotion2vec}, following its use in emotional speech synthesis evaluation~\cite{maestro2025}. Candidate $r$ enters the bank only if the SER model predicts $e_r$ from both $r$ and its CI-simulated version $C(r)$. Accepted references form $\mathcal B$, with $\mathcal B_e\subset\mathcal B$ denoting its emotion-$e$ subset. Although this verifies post-CI recognizability, it does not establish that the expression transfers effectively to every source.

\begin{table*}[t]
\centering
\caption{SER macro-F1 (\%, $\uparrow$) and WER (\%, $\downarrow$) under Clean and PSHC vocoding. Bold: best enhancement system; $^{*}p<.05$ versus Source.}
\label{tab:objective}
\setlength{\tabcolsep}{1.6pt}
\newlength{\resultwidth}
\settowidth{\resultwidth}{\textbf{00.00}}
\newcommand{\rc}[1]{\makebox[\resultwidth][r]{#1}}
\begin{tabular*}{\textwidth}{@{\extracolsep{\fill}}lrrrrrrrrrrrrrrrr@{}}
\toprule
 & \multicolumn{8}{c}{RAVDESS} & \multicolumn{8}{c}{ESD English} \\
\cmidrule(lr){2-9}\cmidrule(lr){10-17}
 & \multicolumn{2}{c}{Clean} & \multicolumn{2}{c}{N4} & \multicolumn{2}{c}{N8} & \multicolumn{2}{c}{N16}
 & \multicolumn{2}{c}{Clean} & \multicolumn{2}{c}{N4} & \multicolumn{2}{c}{N8} & \multicolumn{2}{c}{N16} \\
\cmidrule(lr){2-3}\cmidrule(lr){4-5}\cmidrule(lr){6-7}\cmidrule(lr){8-9}
\cmidrule(lr){10-11}\cmidrule(lr){12-13}\cmidrule(lr){14-15}\cmidrule(lr){16-17}
Model & F1 & WER & F1 & WER & F1 & WER & F1 & WER & F1 & WER & F1 & WER & F1 & WER & F1 & WER \\
\midrule
Source (Normal) & \rc{97.78} & \rc{0.00} & \rc{80.35} & \rc{1.39} & \rc{78.26} & \rc{1.39} & \rc{80.33} & \rc{0.00} & \rc{97.87} & \rc{3.94} & \rc{58.71} & \rc{10.99} & \rc{57.09} & \rc{5.82} & \rc{70.46} & \rc{5.04} \\
Source (Strong) & \rc{95.82} & \rc{0.00} & \rc{75.56} & \rc{3.13} & \rc{83.05} & \rc{1.04} & \rc{89.47} & \rc{0.69} & \rc{--} & \rc{--} & \rc{--} & \rc{--} & \rc{--} & \rc{--} & \rc{--} & \rc{--} \\
\midrule
JD-NMF & \rc{87.26} & \rc{8.33} & \rc{71.17} & \rc{64.58} & \rc{72.26} & \rc{27.43} & \rc{72.84} & \rc{11.81} & \rc{83.46} & \rc{14.64} & \rc{55.90} & \rc{53.10} & \rc{56.20} & \rc{35.91} & \rc{64.04} & \rc{29.51} \\
Fixed reference & \rc{\textbf{100.00}} & \rc{1.74} & \rc{\textbf{84.10}} & \rc{3.82} & \rc{79.23} & \rc{2.43} & \rc{91.77} & \rc{2.08} & \rc{98.44} & \rc{10.42} & \rc{\textbf{70.29}}$^{*}$ & \rc{22.41} & \rc{\textbf{69.66}}$^{*}$ & \rc{14.19} & \rc{\textbf{83.95}}$^{*}$ & \rc{11.98} \\
\midrule
SEER (Proposed) & \rc{\textbf{100.00}} & \rc{\textbf{0.35}} & \rc{80.13} & \rc{\textbf{2.08}} & \rc{\textbf{85.56}} & \rc{\textbf{1.39}} & \rc{\textbf{95.82}}$^{*}$ & \rc{\textbf{1.04}} & \rc{\textbf{99.11}}$^{*}$ & \rc{\textbf{7.13}} & \rc{68.23}$^{*}$ & \rc{\textbf{16.62}} & \rc{68.75}$^{*}$ & \rc{\textbf{10.00}} & \rc{82.06}$^{*}$ & \rc{\textbf{8.48}} \\
\bottomrule
\end{tabular*}
\end{table*}

\subsection{Source-Conditioned Reference Retriever}

Because bank membership does not guarantee successful transfer, SEER embeds each source and reference in a jointly learned latent space and estimates their enhancement utility. Let $E_v\in\mathbb R^{T_v\times1024}$ denote the temporal emotion embedding extracted from utterance $v$ by Maestro-EVC's emotion encoder. For reference $r$, descriptor $d_r$ concatenates the mean and standard deviation of $E_r$; the mean, standard deviation, extrema, and quartiles of smoothed mel-spectral frame energy and mean-normalized $F_0$; voiced-frame ratio; and duration. A reference encoder projects the resulting 2064-D descriptor to $h_r\in\mathbb R^{128}$.

On the source side, a Transformer-based temporal encoder aggregates $E_x$ into $h_x\in\mathbb R^{128}$. An auxiliary classifier produces $\mathbf p_x$; the annotation supervises it and selects $\mathcal B_e$ when available, while its prediction selects the subset otherwise. This task encourages $h_x$ to retain emotion information needed for reference comparison.

Within $\mathcal B_e$, scorer $s_\theta$ maps the absolute difference and element-wise product of the two representations to predicted utility, where a larger value indicates greater expected enhancement benefit:
\begin{equation}
\hat u_\theta(x,r)=s_\theta([|h_x-h_r|;h_x\odot h_r]).
\label{eq:pair-retriever}
\end{equation}
The scorer's 64-D intermediate representation $z_\theta(x,r)$ is reused for exploration. The retriever selects the largest positive score and otherwise retains the source. Because complete utility labels require exhaustive conversion, $u(x,r)$ must instead be learned from sampled pairs.

\subsection{Source-Relative Utility Learning}
\label{sec:reward}

\subsubsection{CI-Aware Guiding Reward}
To obtain this supervision, Emotion2Vec+ Large evaluates each sampled conversion $\tilde x_r$, $r\in\mathcal B_e$, and the matched original $x$ before and after CI simulation. Its nine-class scores are renormalized over Angry, Happy, Sad, and Neutral; $\hat e(v)$ and $p_e(v)$ denote the resulting prediction and intended-emotion score. For $v\in\{x,\tilde x_r\}$, define
\begin{equation}
\begin{aligned}
c(v)&=\mathbb I[\hat e(v)=e\ \land\ \hat e(C(v))=e],\\
q(v)&=\mathbb I[\hat e(v)=e]p_e(C(v)).
\end{aligned}
\label{eq:reward-components}
\end{equation}
The CI-aware guiding reward defines the target utility
\begin{equation}
u(x,r)=c(\tilde x_r)-c(x)+\tau[q(\tilde x_r)-q(x)],\quad \tau=0.1.
\label{eq:utility}
\end{equation}
Thus, $c(\tilde{x}_r)-c(x)$ is $+1$ for repaired retention, $-1$ for harm, and zero otherwise. The $q(v)$ difference adds post-CI confidence when the clean prediction is correct, providing finer guidance toward recognizable expression.

\subsubsection{Reference Exploration}
Because only evaluated pairs yield utility targets, greedy queries can reinforce early errors and concentrate on a few references. We therefore adapt the shallow-exploration principle of Neural-LinUCB~\cite{xu2022neural}. We normalize the 64-D score latent as $\bar z_\theta=z_\theta/\lVert z_\theta\rVert_2$; appending a unit bias and rescaling gives $\psi_\theta=2^{-1/2}[\bar z_\theta;1]\in\mathbb R^{65}$. For $\alpha=1$ and up to 16,384 unique pairs in history $\mathcal H$, the unit-ridge matrix and acquisition score are
\begin{equation}
\begin{aligned}
V&=I_{65}+\!\sum_{(x',r')\in\mathcal H}
\psi_\theta(x',r')\psi_\theta(x',r')^\top,\\
a(x,r)&=\hat u_\theta(x,r)+\alpha
\sqrt{\psi_\theta(x,r)^\top V^{-1}\psi_\theta(x,r)}.
\end{aligned}
\label{eq:ucb}
\end{equation}
The uncertainty bonus favors score latents less covered by $\mathcal H$. For each source, the four highest-scoring references are evaluated with its original-speech baseline. Current outcomes and sampled past pairs form $\mathcal D_t$, while $\mathcal X_t$ denotes the current source batch. The retriever is trained with
\begin{equation}
\begin{aligned}
\mathcal L_{\mathrm{util}}&=\operatorname{MSE}_{\mathcal D_t}
[\hat u_\theta(x,r),u(x,r)],\\
\mathcal L_{\mathrm{emo}}&=\operatorname{CE}_{\mathcal X_t}
[\mathbf p_x,e_x],\qquad
\mathcal L=\mathcal L_{\mathrm{util}}+\mathcal L_{\mathrm{emo}}.
\end{aligned}
\label{eq:retriever-objective}
\end{equation}
$\mathcal L_{\mathrm{util}}$ updates the retriever, while $\mathcal L_{\mathrm{emo}}$ also guides the source encoder. Since $z_\theta(x,r)$ changes during learning, $V$ is rebuilt from $\mathcal H$ every 16 updates. Acquisition is used only for reward evaluation; final retrieval uses $\hat u_\theta$.

\section{Experimental Setup}
\label{sec:experiments}

\subsection{Datasets}
We use RAVDESS~\cite{livingstone2018} (24 actors) and English ESD~\cite{zhou2022esd} (ten speakers), restricted to Angry, Happy, and Sad. Actor-disjoint RAVDESS train/validation/test splits contain 384/96/96 Normal/Strong utterances; ESD has 9,000/600/900. Reference banks built only from training data contain 281 and 6,630 recordings, respectively.

\subsection{Training and Evaluation Setup}
For each dataset, a separate Maestro-EVC backbone trains on the targets plus Neutral and Surprise, while SEER retrieves only the targets. We optimize SEER for N8 by selecting eight maxima from a 22-channel CI24-style PSHC vocoder~\cite{brueggeman2022}; N4/N16 test sparser/denser spectral information~\cite{mo2023}. To expose both outcomes, training batches balance retention/failure, and validation N8 macro-F1 selects the checkpoint. We evaluate emotion with Emotion2Vec+ macro-F1 (Neutral is an error) and intelligibility with Whisper large-v3-turbo WER~\cite{radford2023whisper}. Further implementation details and demo audio are available online.\footnote{\url{https://github.com/HenryChou36/SEER}}

For comparison, Table~\ref{tab:objective} includes Source (Normal), matched RAVDESS Strong, JD-NMF, Fixed reference, and SEER; RAVDESS-trained JD-NMF transfers unchanged to ESD. To test a source-independent choice, Fixed reference selects one template per emotion from SEER's pool at that checkpoint, maximizing the 95\% Wilson lower bound of N8 retention among candidates covering at least 10\% of same-emotion training sources. EVC then uses it for every source of that emotion; retention requires the intended N8 label, and ties are random. All 48 RAVDESS Normal test sources enter objective evaluation and are exhaustively paired with the same-emotion 281-reference bank in Table~\ref{tab:reference-groups}.

For subjective evaluation, 16 listeners each label 279 trials as Angry, Happy, Sad, or Neutral. Trials span 24 repetition-1 sources (four held-out actors, three emotions, two statements), four systems, and N4/N8/N16; nine identity outputs reuse Source judgments. Table~\ref{tab:subjective} reports macro-F1 and Fig.~\ref{fig:listener-confusion} N8 confusions. Within-table Holm correction follows two-sided tests: 10,000 paired permutations at utterance/source levels for Tables~\ref{tab:objective}/\ref{tab:reference-groups} and exact listener sign flips for Table~\ref{tab:subjective}.

\section{Experimental Results and Analysis}
\label{sec:results}

\subsection{Objective Evaluation}

Table~\ref{tab:objective} shows that SEER raises Source macro-F1 at the N8 target by 7.30 points on RAVDESS and 11.66 points on ESD, and exceeds JD-NMF across all six dataset--CI comparisons by 8.96--22.98 points. ESD gains remain significant at N4, N8, and N16, showing that cues learned for N8 transfer to sparser and denser settings. On RAVDESS, the 15.48-point N16 gain is significant, whereas N4 does not improve, indicating less reliable transfer under the sparsest condition.

WER exposes an accompanying intelligibility trade-off, especially with sparse information: at N4, RAVDESS Strong raises WER from 1.39\% to 3.13\%, while Fixed reference raises ESD WER from 10.99\% to 22.41\%. Compared with Fixed reference, SEER raises RAVDESS macro-F1 from 79.23\% to 85.56\% at N8 and from 91.77\% to 95.82\% at N16, while remaining within 0.91--2.05 points on ESD. It also lowers WER in every condition, including 2.43\% to 1.39\% on RAVDESS N8 and 14.19\% to 10.00\% on ESD N8. Source-conditioned retrieval thus limits the intelligibility loss of applying one reference to every source.

\begin{table}[t]
\centering
\caption{Subjective emotion-recognition macro-F1 (\%) under different CI conditions; $^{*}p<.05$ versus Source.}
\label{tab:subjective}
\setlength{\tabcolsep}{2.5pt}
\begin{tabular*}{\columnwidth}{@{\extracolsep{\fill}}lrrr@{}}
\toprule
Method & N4 & N8 & N16 \\
\midrule
Source & 46.42 & 52.66 & 58.71 \\
Strong & 79.90$^{*}$ & 83.11$^{*}$ & 77.91$^{*}$ \\
JD-NMF & 64.06$^{*}$ & 62.97$^{*}$ & 65.71 \\
SEER & 68.37$^{*}$ & 70.14$^{*}$ & 67.39$^{*}$ \\
\bottomrule
\end{tabular*}
\end{table}

\begin{figure}[t]
\centering
\includegraphics[width=\columnwidth]{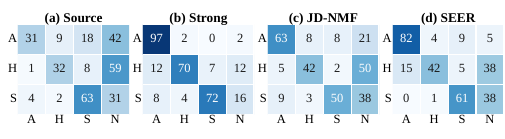}
\caption{N8 subjective confusion matrices (\%), where rows/columns denote intended/perceived emotion and A/H/S/N denote Angry/Happy/Sad/Neutral.}
\label{fig:listener-confusion}
\end{figure}

\subsection{Subjective Evaluation}

Table~\ref{tab:subjective} shows consistent listener benefits: SEER improves Source by 21.94, 17.48, and 8.68 macro-F1 points at N4, N8, and N16, with every gain remaining significant after Holm correction. It also exceeds JD-NMF by 1.68--7.17 points. Strong remains highest at 77.91--83.11\%, confirming that exaggerated expression can be substantially more recognizable under CI processing.

The N8 confusions in Fig.~\ref{fig:listener-confusion} show mixed emotion-dependent effects similar to earlier JD-NMF conversion~\cite{brueggeman2022}. JD-NMF improves Angry and Happy but lowers Sad recognition from 63.3\% to 50.0\%. SEER raises Angry recognition from 31.3\% to 82.0\% and lowers Neutral responses from 42.2\% to 5.5\%. Yet selected references do not yield cues equally distinct from every competing emotion. For Happy, recognition rises from 32.0\% to 42.2\% and Neutral falls from 59.4\% to 37.5\%, but Angry rises from 0.8\% to 14.8\%. For Sad, Angry/Happy responses fall from 5.5\% to 0.8\% but Neutral rises from 31.3\% to 38.3\%, leaving recognition near Source (63.3\% versus 60.9\%). Thus, references improve some emotion boundaries but remain insufficient to separate Happy from Angry or Sad from Neutral after CI processing.

\begin{table}[t]
\centering
\caption{Mean RAVDESS emotion-retention success rates after conversion and CI simulation (\%); $^{*}p<.05$ between the listed groups.}
\label{tab:reference-groups}
\setlength{\tabcolsep}{2.5pt}
\begin{tabular*}{\columnwidth}{@{\extracolsep{\fill}}llrrr@{}}
\toprule
Factor & Groups & N4 & N8 & N16 \\
\midrule
Intensity & Normal/Strong & 56.0/65.1$^{*}$ & 59.4/63.1 & 64.5/75.1 \\
Gender & Diff./Same & 60.7/62.8 & 61.7/62.8 & 70.6/71.3 \\
Content & Diff./Same & 62.0/60.9 & 61.7/61.9 & 69.6/71.4 \\
\bottomrule
\end{tabular*}
\end{table}

\subsection{Successful-Reference Structure}

Table~\ref{tab:reference-groups} shows that same- and different-gender references differ by only 0.7--2.1 points, while same- and different-content references differ by only 0.2--1.8 points; none is significant. Successful conversion therefore does not depend strongly on matching speaker gender or linguistic content in this corpus, supporting a nonparallel reference bank without matched speakers or sentences.

Intensity shows a clearer aggregate pattern: Strong exceeds Normal at N4 (65.1\% versus 56.0\%), N8 (63.1\% versus 59.4\%), and N16 (75.1\% versus 64.5\%). Only N4 remains significant after Holm correction, providing the strongest evidence under sparse processing. Strong expression is therefore useful in aggregate, but its nonsignificant N8/N16 differences leave intensity insufficient as a source-level rule and motivate pairwise utility estimation.
\section{Conclusion}

SEER learns CI-aware source--reference utility from sampled conversions without exhaustive pairing. This enables source-conditioned enhancement from categorical corpora without parallel recordings or intensity labels. Objective and listener results show improved post-CI emotion recognition with less intelligibility loss than forced-reference conversion. Strong expression helps overall, but matched gender/content and intensity alone do not determine utility. Future work will expand emotion and simulator coverage and add CI-listener rewards.

\section{Acknowledgment}
OpenAI ChatGPT was used for language and style editing and to assist in refining
the final delivered code. The authors developed and conducted all technical
content, experimental design, analyses, and conclusions, and take full
responsibility for the manuscript and accompanying code.

\bibliographystyle{IEEEbib}
\bibliography{refs}
\end{document}